\documentclass[12pt]{article}

\usepackage{setspace}
\usepackage{graphicx}
\usepackage{enumerate}
\usepackage{float}
\usepackage[top=1.5in, bottom=1.5in, left=1in, right=1in]{geometry}
\usepackage{hyperref}
\usepackage{rotating}
\usepackage{color}
\usepackage{amsmath}
\usepackage{verbatim}
\usepackage{cleveref}
\usepackage{pslatex}
\usepackage{caption}
\usepackage{subcaption}
\usepackage{placeins}
\usepackage[T1]{fontenc}
\newcommand{\ONOOHlong}{14}
\newcommand{\OHfromONOOH}{15}
\newcommand{\ONOOHshort}{16}
\newcommand{\NitrousAcidAssociation}{19}
\newcommand{\WaterAssociation}{20}

\title{Momentum, Heat, and Neutral Mass Transport in Convective Atmospheric Pressure Plasma-Liquid Systems and Implications for Aqueous Targets}
\author{Alexander Lindsay, Carly Anderson, Elmar Slikboer, Steven Shannon, and David Graves}

\begin{document}
\maketitle

\begin{abstract}

There is a growing interest in the study of plasma-liquid interactions with application to biomedicine, chemical disinfection, agriculture, and other fields. This work models the momentum, heat, and neutral species mass transfer between gas and aqueous phases in the context of a streamer discharge; the qualitative conclusions are generally applicable to plasma-liquid systems. The problem domain is discretized using the finite element method. The most interesting and relevant model result for application purposes is the steep gradients in reactive species at the interface. At the center of where the reactive gas stream impinges on the water surface, the aqueous concentrations of OH and ONOOH decrease by roughly 9 and 4 orders of magnitude respectively within 50 $\mu$m of the interface. Recognizing the limited penetration of reactive plasma species into the aqueous phase is critical to discussions about the therapeutic mechanisms for direct plasma treatment of  biological solutions. Other interesting results from this study include the presence of a 10 K temperature drop in the gas boundary layer adjacent to the interface that arises from convective cooling and water evaporation. Accounting for the resulting difference between gas and liquid bulk temperatures has a significant impact on reaction kinetics; factor of two changes in terminal aqueous species concentrations like H$_2$O$_2$, NO$_2^-$, and NO$_3^-$ are observed if the effect of evaporative cooling is not included.
    
\end{abstract}

\section{Introduction}

There is a general interest in the study of plasma-liquid interactions within the
scientific community for an array of applications, including but not limited to biomedicine and biological disinfection \cite{Kong2009b,Laroussi2009,Shimizu2014c,VonWoedtke2014a,VonWoedtke2013a,Joubert2013a}, chemical disinfection \cite{Johnson2006,Locke2006,Theron2008}, and agricultural applications. \cite{Park2013b,Lindsay2014} In order to successfully realize these applications and develop mature technologies, the basic science underlying these coupled plasma gas-liquid systems must be well understood. Recent experimental \cite{Lukes2014b,Bruggeman2009d,Pavlovich2013g,Traylor2011h} and modeling work \cite{Babaeva2014b,Tian2014,Chen2014a} have revealed many of the important physiochemical processes. However, more work remains, especially in the understanding of multi-phase convective systems like jets or streamers over water.

This paper addresses several points regarding transport in plasma-liquid systems. First, convection-induced evaporation of water can lead to rather large temperature gradients at the gas-liquid interface, as large as 8 Kelvin/100 microns in the gas boundary layer. The drop in temperature from gas to liquid due to evaporative cooling has not yet been considered in the plasma literature; it is directly relevant to plasma chemistry because of the Arrhenius dependence of many reaction rate constants on temperature. In addition to evaporative cooling, convection leads to depletion of water vapor concentration in the vicinity of the discharge. This will also impact gas-phase plasma chemical reactions that depend on water vapor as a reactant. \cite{winter2013feed}

This study illustrates the sharp drop in the concentrations of highly reactive species from the liquid surface to the liquid bulk. In the liquid near the interface, OH concentrations fall by as many as 9 orders of magnitude over tens of microns. Though the existence of strong gradients in aqueous radical concentrations is well known in the biochemistry literature, \cite{Bachi2013,Halliwell}, it has yet to receive significant attention in the plasma-liquid and plasma-medicine communities. Reference \cite{Chen2014a} explicitly models the limited penetration of primary plasma-generated species into aqueous solutions. Recognizing these strong gradients in reactive radical concentrations has important implications for understanding plasma therapeutics in which cellular effects are often observed a significant distance (1 cm in case of subcutaneous tumors \cite{Graves2014review}) away from local plasma treatment. The results shown here are consistent with the model developed by Graves \cite{Graves2014review} where short-lived plasma-generated radicals from the gas phase react in the surface layer of the aqueous/biological phase to form longer-lived species such as oxidized and/or nitrated/nitrosylated proteins, peptides, lipids which can then enter or communicate with surface cells.      

\section{Model Description}

\begin{figure}[htpb]
    \centering
        \includegraphics[width=.5\textwidth]{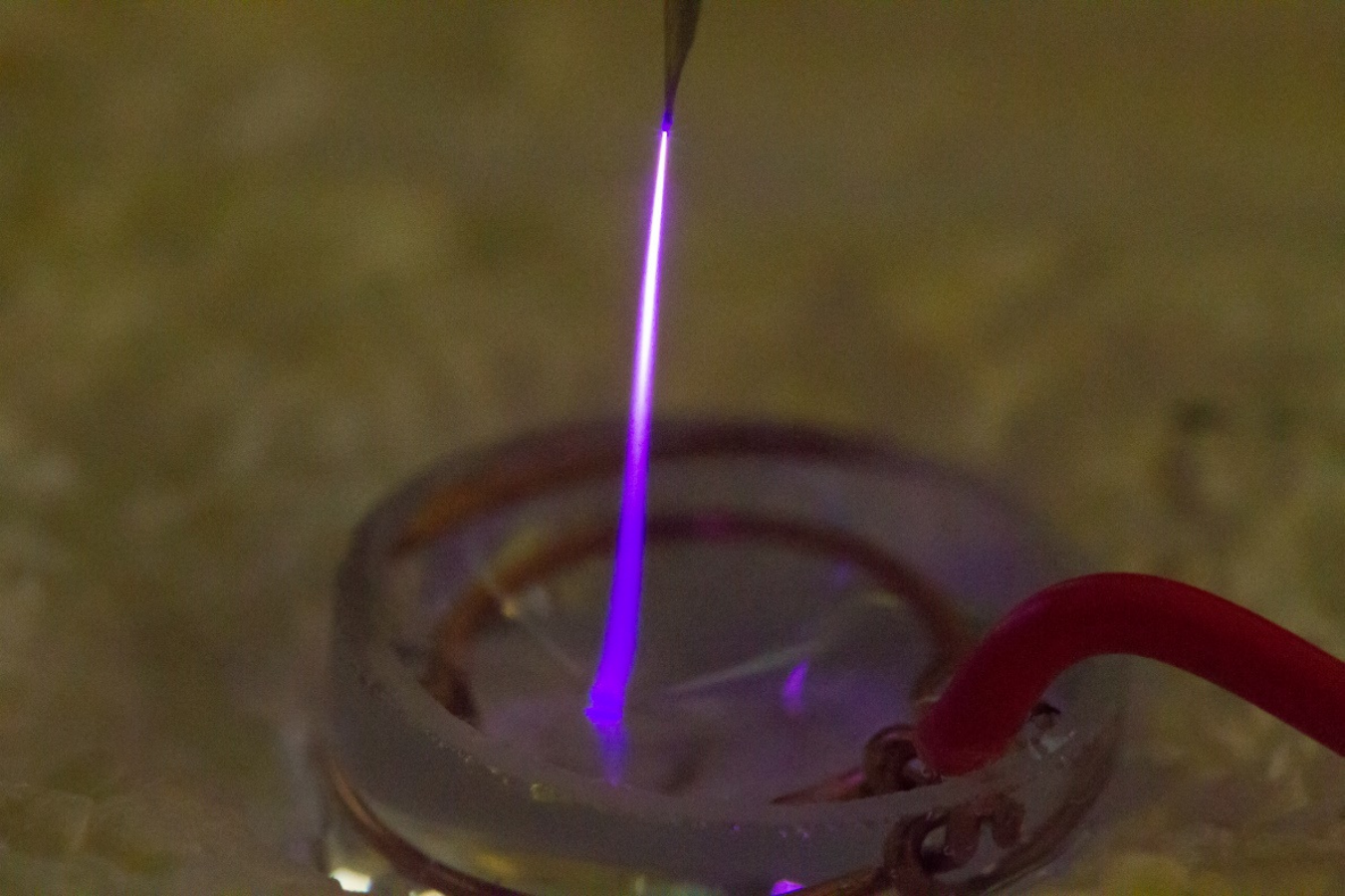}
    \caption{Experimental set-up for pulsed streamer-liquid system. Discharge voltages are typically between 6 and 8 kV with pulse frequencies between 10 and 30 kHz. The gap from the needle tip to the water surface is between 3 and 15 mm. Water treatment volumes are between 10 and 20 mL.}
    \label{fig:streamer_picture}
\end{figure}

The convective system chosen for modelling is shown in \cref{fig:streamer_picture}. It is essentially a point-to-plane pulsed-streamer in which the liquid surface serves as the gas discharge cathode. The streamer is self-pulsed because of a ballasting resistor; typical discharge voltages are between 6 and 8 kV and pulse frequencies are generally 10 to 30 kHz. The sharp anode tip can be stationed anywhere between 3 and 15 mm above the water surface. The water is contained within a glass petri dish of radius 3 cm; water treatment volumes are generally between 10 and 20 mL.  

We have chosen not to explicitly model the atmospheric discharge due to the computational cost of modeling of gas-liquid interfacial transport over time-scales of minutes (the equivalent of millions of individual pulses). Instead, gas phase concentrations of the species modeled are taken from the DBD-water calculations in \cite{Tian2014} and diluted by a factor of 10; the gas phase species included and their respective inlet concentrations are shown in table \ref{tab:inletconc}. The factor 10 dilution is done for numerical stability.  High concentrations of reactive oxygen and nitrogen species (RONS) at the interface combined with very fast rate coefficients, particularly for reactions involving OH, can lead to singularities over our time scales of interest unless the computational mesh is sufficiently refined, requiring significant computational power. All concentration results and discussion will be focused on relative magnitudes and variation with space and time. The results in \cite{Tian2014} are chosen as an input basis because of the absence of experimental work for atmospheric plasmas in which a suite of gas-phase plasma species concentrations are reported. Most of the measurements in the literature focus on single species like OH \cite{ono1998measurement,ono2001oh,nakagawa2011density,verreycken2012time}, atomic oxygen \cite{niemi2005absolute}, or NO \cite{kanazawa2003two} that are accessible through Laser Induced Fluorescence (LIF).

A key feature of streamer discharges is the ionic wind. The ionic wind is a net flow of gas towards the cathode that results from the effective drag of cations on neutral gas molecules. In the streamer-liquid system studied, the ionic wind creates convective forces that affect species transport at the gas-liquid interface. Since the discharge is not modelled, another mechanism must be used to produce the ionic wind present in a streamer discharge. The mechanism chosen is a jet-like inlet with diameter equal to the diameter of the needle used in the experimental streamer set-up. The velocity profile at the inlet of the jet channel is chosen such that the channel exit profile closely mimics that expected in a streamer experiment. The profile is built from the expected streamer maximum axial velocity which is in turn determined from the modeling work of Zhao et. al. \cite{Zhao2005a} Using values of 6 kV for the discharge voltage and 6.5 mm for the gap distance, the maximum axial velocity is interpolated to be 7.75 m/s near the needle tip. The spatial inlet profile is then computed from \cref{eq:v_in} assuming laminar no-slip conditions in the inlet channel: \cite{bird2007transport}

\begin{equation}
    v_{z} = v_{z,max}\left(1-\left(\frac{r}{R}\right)^2\right)
    \label{eq:v_in}
\end{equation}

\begin{figure}[htb]
    \centering
        \includegraphics[width=\textwidth]{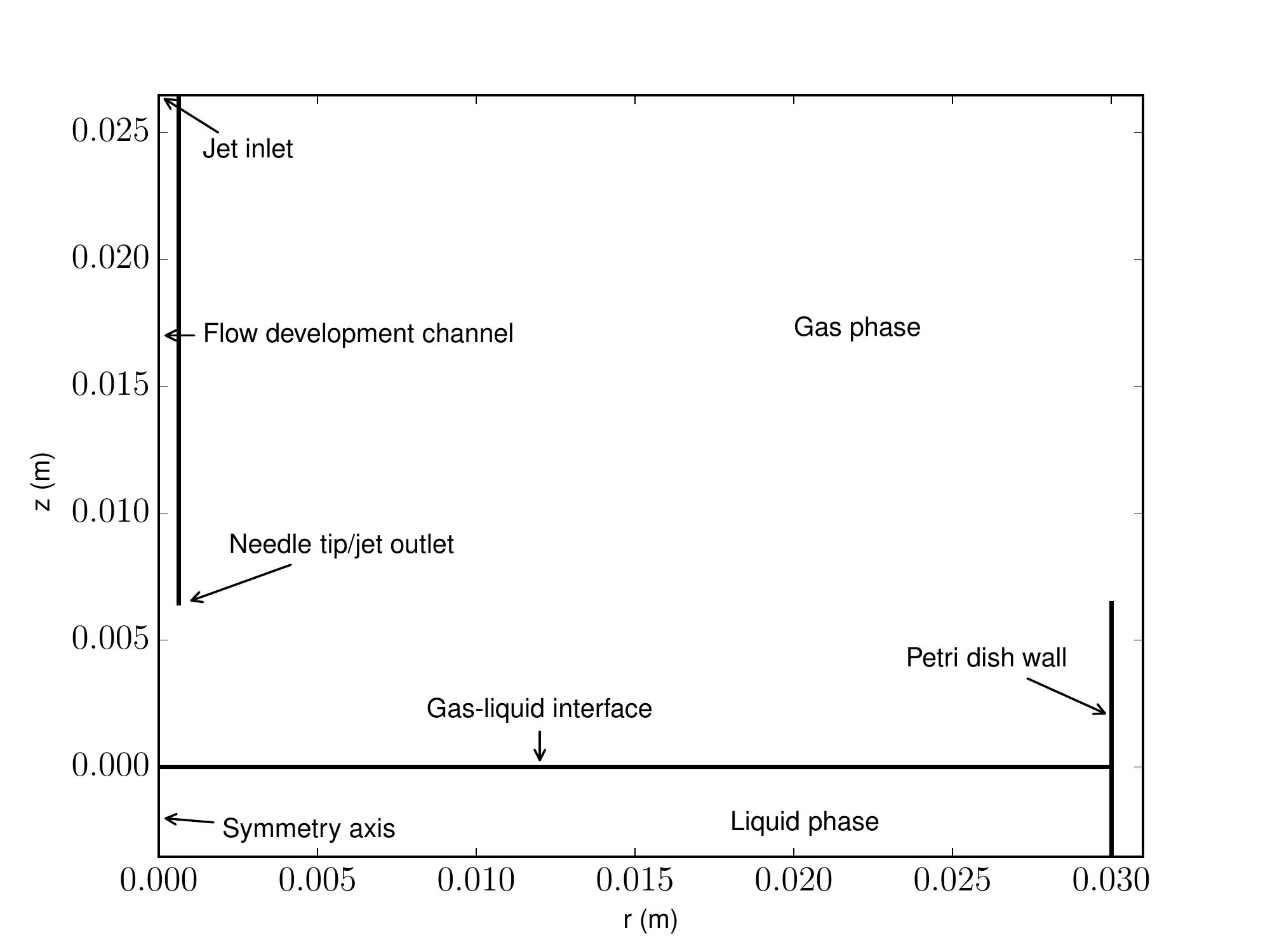}
    \caption{Experimental set-up for pulsed streamer-liquid system. Axis units are meters. The Python script used to create this figure, as well as the scripts used to create all subsequent figures can be found at \cite{scriptsLocation}}
    \label{fig:model_geom}
\end{figure}

 The model geometry is shown in \cref{fig:model_geom}; model inputs are summarized in \cref{tab:species_list,tab:gen_inputs,tab:henryconstants,tab:diffusioncoef,tab:inletconc}. The governing equations used to model the gas and liquid phases are the incompressible Navier-Stokes equations for momentum transport and convection-diffusion equations for heat and mass transport; they are shown in \cref{eq:mass_cont,eq:nav_stokes,eq:heat,eq:mass}

\begin{equation}
    \nabla\cdot\vec{u}=0
    \label{eq:mass_cont}
\end{equation}
\begin{equation}    
    \rho \left(\frac{\partial\vec{u}}{\partial t}+\vec{u}\cdot\nabla\vec{u}\right)=-\nabla p+\mu \nabla^2 \vec{u}
    \label{eq:nav_stokes}
\end{equation}
\begin{equation}
    \rho C_p\vec{u}\cdot\nabla T = \nabla\cdot\left(k\nabla T\right)
    \label{eq:heat}
\end{equation}
\begin{equation}
    \nabla\cdot\left(-D_i\nabla C_i\right) + \vec{u}\cdot\nabla C_i = R_i
    \label{eq:mass}
\end{equation}

with $\vec{u}$ representing the fluid velocity, $\rho$ the overall mass density, p the static pressure, $\mu$ the dynamic viscosity, C$_p$ the constant pressure heat capacity, T the temperature, k the thermal conductivity, D$_i$ the diffusivity of species i, C$_i$ the concentration of species i, and R$_i$ the source term representing chemical reactions. When solving the model equations, the transient equations \ref{eq:mass_cont} and \ref{eq:nav_stokes} are solved until the velocity components and pressure reach a steady-state. Then the steady-state velocity field is inserted into the convection terms in equations \ref{eq:heat} and \ref{eq:mass}. The heat and mass transport equations are typically solved for a physical time of 1000 seconds in order to match common experimental treatment times in our laboratory. The temperature dependence of transport parameters in equations \ref{eq:heat} and \ref{eq:mass}, including reaction rate coefficients, is considered. Gaseous diffusion coefficients are assumed to scale with T$^{3/2}$ as predicted by Chapman-Enskog theory. \cite[p. 119]{cussler2009diffusion} The temperature dependence of liquid phase diffusion coefficients is constructed using the Stokes-Einstein equation \cite[p. 529]{bird2007transport} and the following equation for viscosity: \cite[p. 31]{bird2007transport} 

\begin{equation}
	\mu \propto exp(3.8T_b/T)
	\label{eq:viscosityTempDependence}
\end{equation}

where T$_b$ is the boiling point of the solvent (373 K in the case of water). Reaction rate coefficients and their temperature dependence are contained in table \ref{tab:rxns}. Because the fluid flow equations are solved prior to solution of the heat transport equation, the temperature dependence of transport parameters in equations \ref{eq:mass_cont} and \ref{eq:nav_stokes} is not included in this work. Model boundary conditions are as follows: for fluid flow, all solid walls are assumed to be no-slip, i.e. all velocity components at the walls are zero. For most results presented the gas-liquid interface is assumed to be flat and static. For the static interface, the z-velocity component at the interface is set to zero in both the gas and liquid phases. Both the normal and shear stresses are continous across the interface. One may question whether the assumption of a flat interface is valid when in reality the interface is deformed by the gas flow. In the results section it will be demonstrated that for the purpose of this work, an assumption of a flat interfaction is sufficient. For temperature calculations, all solid surfaces are assumed to be insulated; at the interface, the temperature is assumed continuous. Additionally, evaporation of water at the surface is coupled to the system's heat transport in the following way: \cite{bird2007transport}

\begin{equation}
    Q_b = J_{z,H_2O}\cdot H_{vap} = -D_{H_2O,g}\cdot \left.\frac{\partial C_{H_2O(g)}}{\partial z}\right|_{z=interface}\cdot H_{vap}
    \label{eq:vapheatsrc}
\end{equation}

where Q$_b$ is the heat flux, J$_{z,H_2O}$ is the molar flux of H$_2$O coming from evaporation, and H$_{vap}$ is the latent heat of vaporization for water. The concentration of water vapor at the interface is determined from Antoine's equation: \cite{antoine1888thermodynamic}

\begin{equation}
    \log_{10}p = 8.07131 - \frac{1730.63}{233.426+T}
    \label{eq:Antoine}
\end{equation}

with p in units of mmHg, T in units of degrees Celsius, and the constant values taken from \cite{dimian2014integrated}.  The temperature at the gas inlet is set to 300 K for all times. Initially the temperature in both gas and liquid domains is 300 K. Inlet concentrations for species other than water vapor are specified at the beginnning of the jet flow development channel and are given in \cref{tab:inletconc}. Dilute species that are present in both gas and aqueous phases (OH, H$_2$O$_2$, NO, NO$_2$, N$_2$O$_4$, HNO$_2$, and HNO$_3$) have continuous fluxes across the gas-liquid interface. Concentrations immediately above and below the interface are assumed to be in equilibrium as described by their Henry's law coefficients, listed in \cref{tab:henryconstants}. A limited set of gas and liquid phase reactions encompassing the most important NO$_x$ chemistry is used to reduce computational expense. The reactions used and their corresponding rate coefficients are listed in \cref{tab:rxns}. It should be noted that because of its highly acidic nature (pKa ~ -1.3), HNO$_3$ is assumed to dissociate into H$^+$ and NO$_3^-$ immediately after entering the aqueous phase. The model equations are solved using the finite element method implemented in Comsol Multiphysics version 4.4. Copies of the model are freely available upon request.

When the ``discharge region" is discussed, it is in reference to the gas region roughly demarcated in the z direction by the jet outlet/needle anode and the water surface, and in the radial direction by the radius of the jet channel/needle, where the plasma is visible during experiments. The reader is reminded that the plasma and its electrodynamics are not explicitly modeled; the focus of this investigation is the qualitative behavior of momentum, heat, and neutral species mass transport in convective systems. The qualitative conclusions drawn here are equally applicable to atmospheric jets or streamer systems in which the ionic wind plays a key role in momentum transport. Moreover, we postulate that the steep gradients in highly reactive neutral species concentrations at the gas-liquid interface shown in the proceeding section are a universal phenomena of atmospheric plasma-liquid systems, regardless of whether they are convective or diffusive; this is consistent with the recent research by \cite{Chen2014a} as well as the greater reservoir of biochemistry literature \cite{Halliwell}. Morever, these gradients have important implications for plasma medicine, mainly that cellular responses must be induced through secondary as opposed to directly generated plasma reactivity.  

\begin{table}[htpb]
    \begin{center}
        \begin{tabular}{|l| p{10cm}|}
            \hline
            Gas phase species & OH, H$_2$O$_2$, NO, NO$_2$, N$_2$O$_4$, HNO$_2$, HNO$_3$, H$_2$O \\
            \hline
            Liquid phase species & OH, H$_2$O$_2$, NO, NO$_2$, N$_2$O$_4$, HNO$_2$, NO$_2^-$, NO$_3^-$, ONOOH, H$^+$, OH$^-$ \\
            \hline
        \end{tabular}
    \end{center}
    \caption{Species included in the model}
    \label{tab:species_list}
\end{table}

\begin{table}[htpb]
    \begin{center}
        \begin{tabular}{|c |c |}
        \hline
        Needle diamater & 1.2 mm \\
        \hline
        Petri dish diameter & 6 cm \\
        \hline
        Gap distance & 6.5 mm \\
        \hline
        Water volume & 10 mL \\
        \hline
        Jet channel length & 2 cm \\
        \hline
        Maximum axial velocity & 7.75 m/s \\
        \hline
        \end{tabular}
    \end{center}
    \caption{General model inputs}
    \label{tab:gen_inputs}
\end{table}
       
    \begin{table}[htpb] 
        \begin{center}
            \begin{tabular}{c |c }\rmfamily
                Molecule & H$_i$ [unitless] \\ \hline \hline
                OH & $6.92\cdot10^2$\\
                H$_2$O$_2$ & $1.92\cdot10^6$\\
                NO & $4.4\cdot10^{-2}$\\
                NO$_2$ & $2.8\cdot10^{-1}$\\
                N$_2$O$_4$ & $3.69\cdot10^1$\\
                HNO$_2$ & $1.15\cdot10^3$\\
                HNO$_3$ & $4.8\cdot10^6$\\
                HOONO & $4.8\cdot10^6$\\
            \end{tabular}
    \end{center}
        \caption{The Henry's constant for a number of molecules\cite{Tian2014}.}
        \label{tab:henryconstants}
    \end{table}


    \begin{table}[htpb] 
        \begin{center}
            \begin{tabular}{c |c |c}\rmfamily
                Molecule & D [m$^2$ s$^{-1}$]& reference\\ \hline \hline
                OH(g) & $4\cdot10^{-5}$ & \cite{Sakiyama2012b}\\
                H$_2$O$_2$(g) & $2\cdot10^{-5}$ & \cite{Sakiyama2012b}\\
                NO(g) & $2\cdot10^{-5}$ & \cite{Sakiyama2012b}\\
                NO$_2$(g) & $1.7\cdot10^{-5}$ & \cite{Sakiyama2012b}\\
                N$_2$O$_4$(g) & $1\cdot10^{-5}$ & \cite{Sakiyama2012b}\\
                HNO$_2$(g) & $2.1\cdot10^{-5}$ & \cite{Sakiyama2012b}\\
                HNO$_3$(g) & $2.1\cdot10^{-5}$ & \cite{Sakiyama2012b}\\
                H$_2$O(g) & $2.3\cdot10^{-5}$ & \cite{Sakiyama2012b}\\
                OH(aq) & $2\cdot10^{-9}$ & Estimate\\
                H$_2$O$_2$(aq) & $1.7\cdot10^{-9}$ & \cite{mcmurtrie1948measurement}\\
                NO(aq) & $2.2\cdot10^{-9}$ & \cite{zacharia2005diffusivity}\\
                NO$_2$(aq) & $2\cdot10^{-9}$ & \cite{skinn2013nitrogen}\\
                N$_2$O$_4$(aq) & $1.5\cdot10^{-9}$ & Estimate\\
                HNO$_2$(aq) & $2.5\cdot10^{-9}$ & By analogy with nitric acid\\
                HNO$_3$(aq) & $2.5\cdot10^{-9}$ & \cite{wills1971diffusion}\\
                ONOOH(aq) & $2.5\cdot10^{-9}$ & By analogy with nitric acid\\    
                NO$^{-}_2$(aq) & $1.7\cdot10^{-9}$ & \cite{kreft2001individual}\\
                NO$^{-}_3$(aq) & $1.7\cdot10^{-9}$ & \cite{kreft2001individual}\\
                H$^{+}$(aq) & $7\cdot10^{-9}$ & \cite{agmon1995grotthuss}\\
                OH$^{-}$(aq) & $1.7\cdot10^{-9}$ & By anology with other anions\\
            \end{tabular}
        \end{center}
        \caption{The diffusion coefficients for a number of molecules at 300 K. See text for implementation of temperature dependence.}
        \label{tab:diffusioncoef}    
    \end{table}

\begin{table}[htpb]
    \begin{center}
        \begin{tabular}{c |c}\rmfamily
           Molecule & Molecule inlet concentration [m$^{-3}$]\\ \hline \hline        
            OH & $1.3\cdot10^{18}$ \\
            H$_2$O$_2$ & $1.6\cdot10^{17}$ \\    
            NO & $8\cdot10^{18}$ \\
            NO$_2$ & $5\cdot10^{16}$ \\
            N$_2$O$_4$ & 0 \\
            HNO$_2$ & $8\cdot10^{17}$ \\
            HNO$_3$ & $9\cdot10^{16}$ \\
            H$_2$O & 0 \\
        \end{tabular}
    \end{center}
    \caption{Gaseous species inlet concentrations. \cite{Tian2014} See text for discussion}
    \label{tab:inletconc}
\end{table}

\begin{table}[htpb]
    \begin{center}
        \begin{tabular}{l |p{6cm} |p{3cm}}\rmfamily
            Reaction & Rate coefficient (Units of s$^{-1}$, m$^3$ mol$^{-1}$ s$^{-1}$, or m$^6$ mol$^{-2}$ s$^{-1}$. Temperature in K. Concentration of M in gas and H$_2$O in liquid are lumped into rate coefficient) & Reference \\ \hline \hline
            Gas phase reactions \\
            1. 2NO$_2 \rightarrow$ N$_2$O$_4$ & $6.02\cdot10^5\cdot(300/T)^{3.8}$ & \cite{Sakiyama2012b} \\
            2. N$_2$O$_4 \rightarrow$ 2NO$_2$ & $4.4\cdot10^6\cdot\exp(-4952/T)$ & \cite{Sakiyama2012b} \\
            3. NO + OH + M $\rightarrow$ HNO$_2$ + M & $1.1\cdot10^{7}\cdot(300/T)^{2.4}$ & \cite{Sakiyama2012b} \\
            4. NO + NO$_2$ + H$_2$O $\rightarrow$ 2HNO$_2$ & 22 & \cite{wayne1951kinetics} \\
            5. 2OH + M $\rightarrow$ H$_2$O$_2$ + M & $1.0\cdot10^7\cdot(T/300)^{-0.8}$ & \cite{Sakiyama2012b} \\
            6. NO$_2$ + OH + M $\rightarrow$ HNO$_3$ + M & $3.2\cdot10^7\cdot(300/T)^{2.9}$ & \cite{Sakiyama2012b} \\
            7. OH + HNO$_2$ $\rightarrow$ NO$_2$ + H$_2$O & $3.0\cdot10^6\cdot\exp(-390/T)$ & \cite{Sakiyama2012b} \\
            8. 2HNO$_2$ $\rightarrow$ NO + NO$_2$ + H$_2$O & $6.0\cdot10^{-3}$ & \cite{Sakiyama2012b} \\
            9. HNO$_2$ + HNO$_3$ $\rightarrow$ 2NO$_2$ + H$_2$O & 9.6 & \cite{Sakiyama2012b} \\
            10. HNO$_3$ + NO $\rightarrow$ HNO$_2$ + NO$_2$ & $4.4\cdot10^{-3}$ & \cite{dorai2002modeling} \\
            \\
            Liquid phase reactions \\
            11. N$_2$O$_4$ + H$_2$O $\rightarrow$ NO$_2^-$ + NO$_3^-$ + 2H$^+$ & 1000 & \cite{coddington1999hydroxyl} \\
            12. 2NO$_2$ + H$_2$O $\rightarrow$ NO$_2^-$ + NO$_3^-$ + 2H$^+$ & $8.4\cdot10^{4}\cdot\exp(-0.033\cdot(T-295))$ & \cite{park1988solubility} \\
            13. NO + NO$_2$ + H$_2$O $\rightarrow$ 2NO$_2^-$ + 2H$^+$ & $1.6\cdot10^{5}$ & \cite{park1988solubility} \\
            \ONOOHlong{}. NO$_2^-$ + H$_2$O$_2$ + H$^+$ $\rightarrow$ ONOOH + H$_2$O & $1.1\cdot10^{-3}$ & \cite{Lukes2014b}\\
            \OHfromONOOH{}. ONOOH $\rightarrow$ .7(NO$_3^-$ + H$^+$) + .3(NO$_2$ + OH) & .8 & \cite{coddington1999hydroxyl}\\
            \ONOOHshort{}. NO$_2$ + OH $\rightarrow$ ONOOH & $5.3\cdot10^{6}$ & \cite{Lukes2014b,goldstein2005chemistry}\\
            17. 2OH + M $\rightarrow$ H$_2$O$_2$ + M & $1\cdot10^{7}\cdot\exp(-450\cdot(1/T-1/298))$ & \cite{johnaelliot1990estimation}\\
            18. HNO$_2$ $\rightarrow$ H$^+$ + NO$_2^-$ & $3.51\cdot10^{6}$ & From reaction \NitrousAcidAssociation{} and equilibrium constant \\
            \NitrousAcidAssociation{}. H$^+$ + NO$_2^-$ $\rightarrow$ HNO$_2$ & $7\cdot10^{6}$ & Assumes diffusion limited\\
            \WaterAssociation{}. H$^+$ + OH$^-$ $\rightarrow$ H$_2$O & $7\cdot10^{6}$ & Assumes diffusion limited \\
            21. H$_2$O $\rightarrow$ H$^+$ + OH$^-$ & $7\cdot10^{-2}$ & From reaction \WaterAssociation{} and equilibrium constant\\
            22. NO + OH $\rightarrow$ HNO$_2$ & $2\cdot10^{7}$ & \cite{Tian2014}\\
			23. 2HNO$_2$ $\rightarrow$ NO + NO$_2$ & $1.34\cdot10^{-2}\cdot\exp(.106\cdot(T-295))$ & \cite{park1988solubility}
        \end{tabular}
    \end{center}
    \caption{Reactions considered in model}
    \label{tab:rxns}
\end{table}

\section{Results and Discussion}

\begin{figure}[htb]
    \centering
        \includegraphics[width=\textwidth]{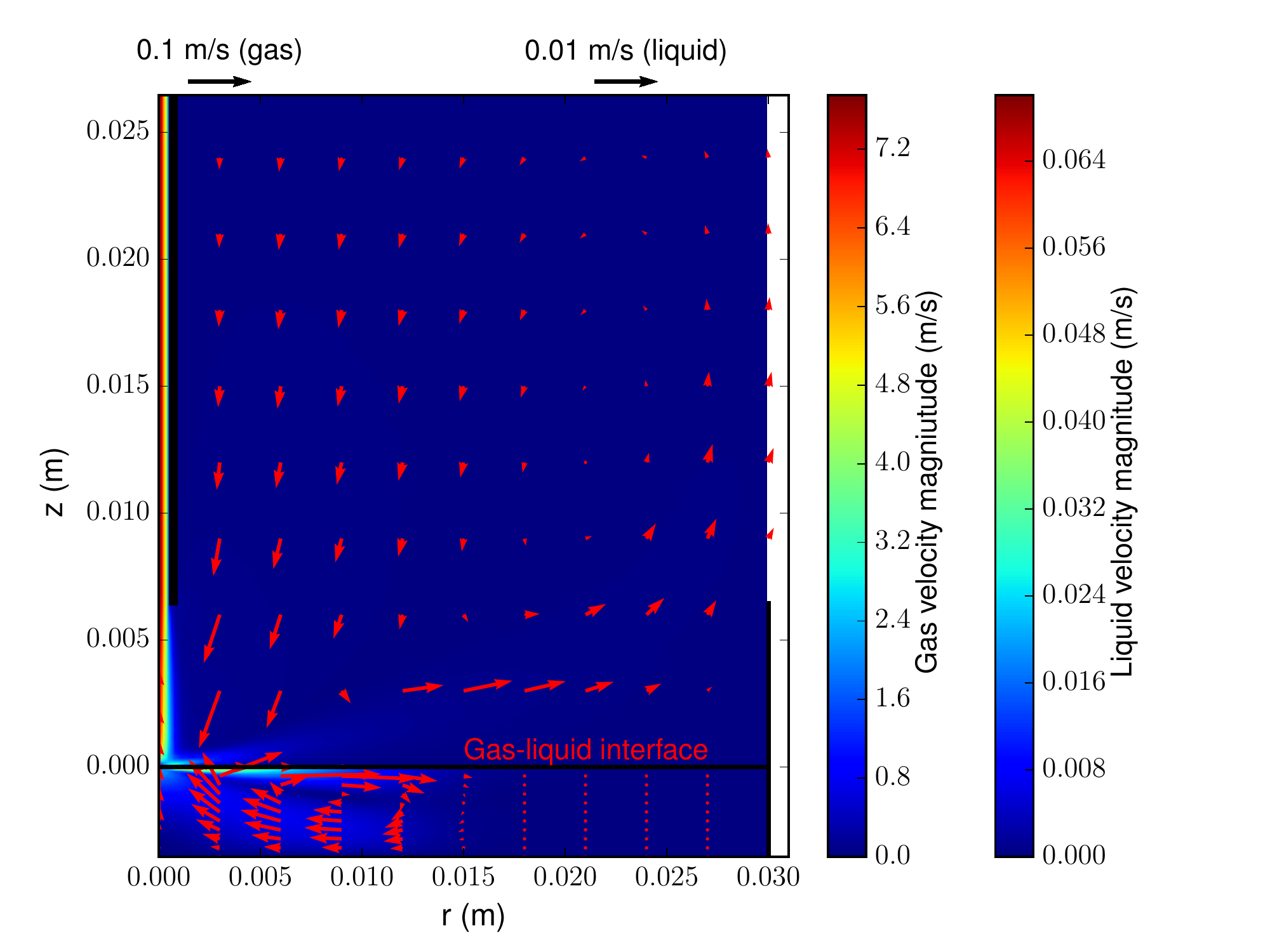}
    \caption{Velocity magnitude and direction. Axis units are meters. Interface is at z = 0. Axis of symmetry is at r = 0. Selected velocity vectors are overlaid on color scale indicating velocity magnitude. Red color represents higher values and blue color represents lower values. Velocity vector arrows are scaled 12.5 times larger in the aqueous phase.}
    \label{fig:v_field}
\end{figure}

The steady state velocity field is shown in figure \ref{fig:v_field}. The recirculating pattern observed in the gas phase is consistent with the corona discharge modelling results reported in \cite{Zhao2005a}. A similar recirculation pattern is observed in the liquid phase, induced by shear stresses between the phases at the gas-liquid interface. The maximum liquid phase velocity, .1 m/s, occurs along the interface approximately .8 mm away from the stagnation point (r = 0).   

\begin{figure}[htb]
    \centering
        \includegraphics[width=\textwidth]{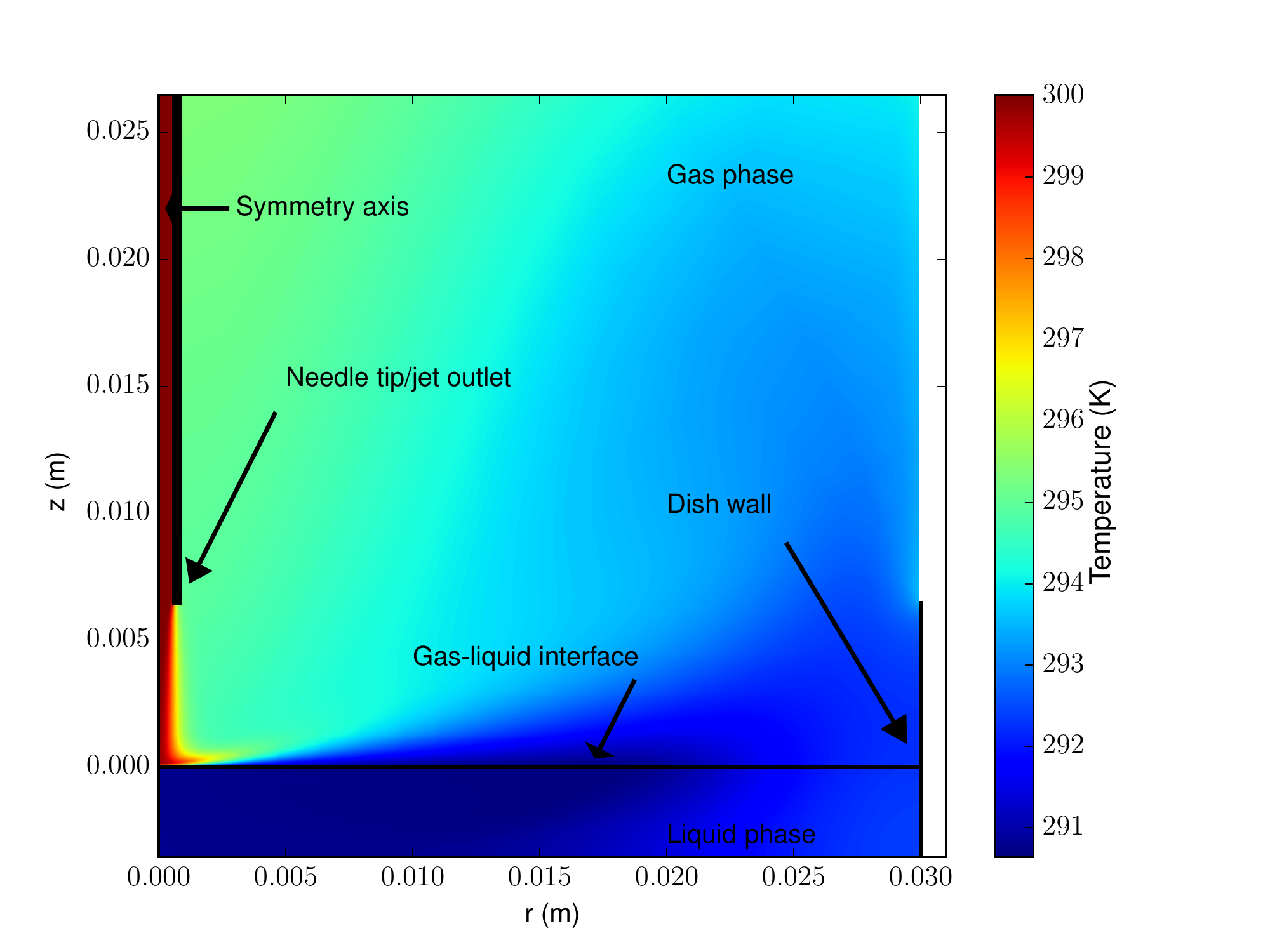}
        \caption{2D temperature profile at t = 1000 seconds. Red color represents higher values; blue color lower values. Inlet temperature is 300 K. Temperature in the bulk liquid has cooled by approximately 10 K because of convection-induced evaporative cooling.}
        \label{fig:temp_2D_profile}
\end{figure}

\begin{figure}[htb]
    \centering
        \includegraphics[width=\textwidth]{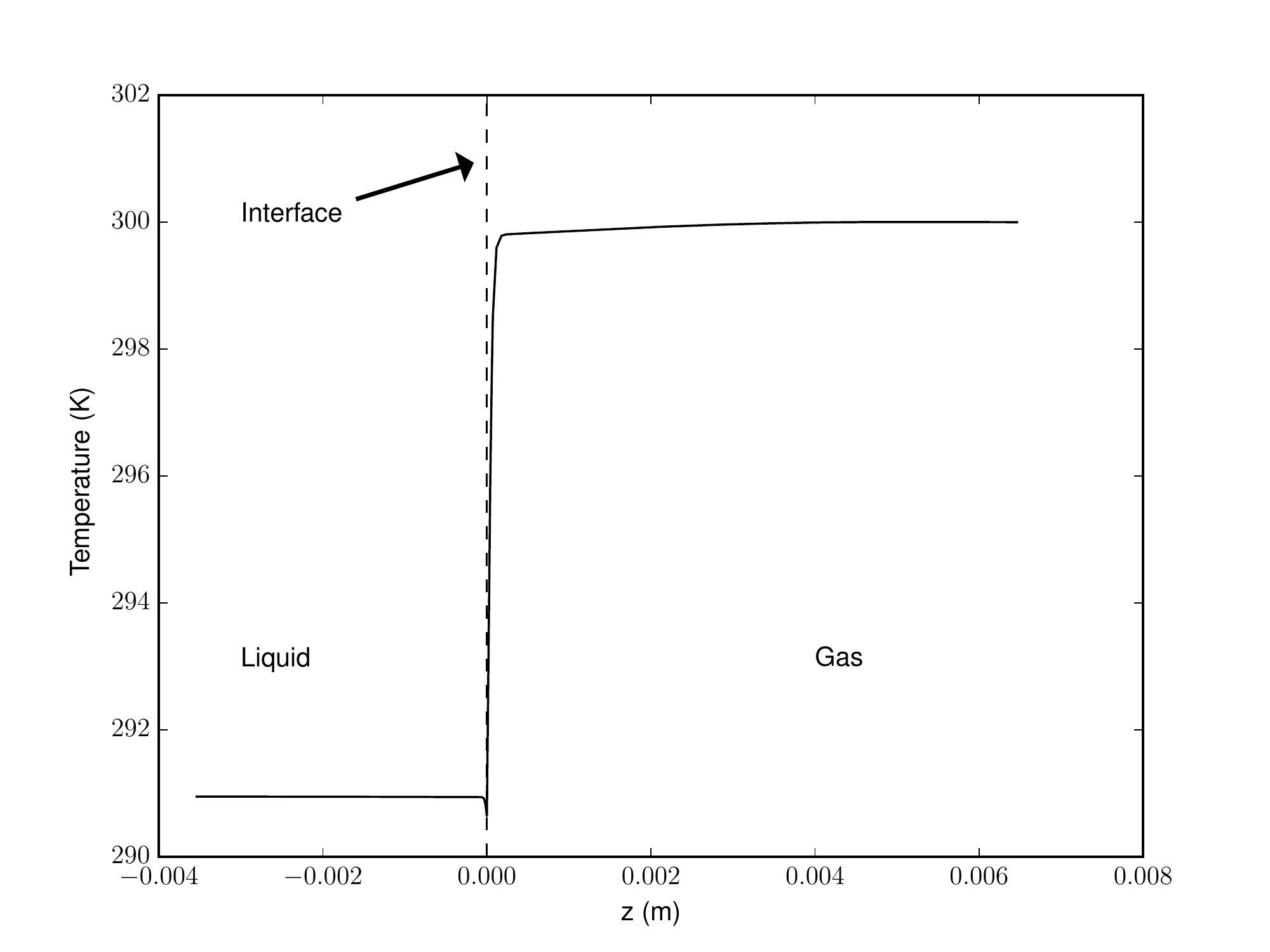}
        \caption{Temperature along z-axis. Illustrates the large temperature gradient that exists at the gas-liquid interface and the resulting difference in bulk gas and liquid temperatures.}
        \label{fig:temp_1D_profile}
\end{figure}

\begin{figure}[htb]
    \centering
        \includegraphics[width=\textwidth]{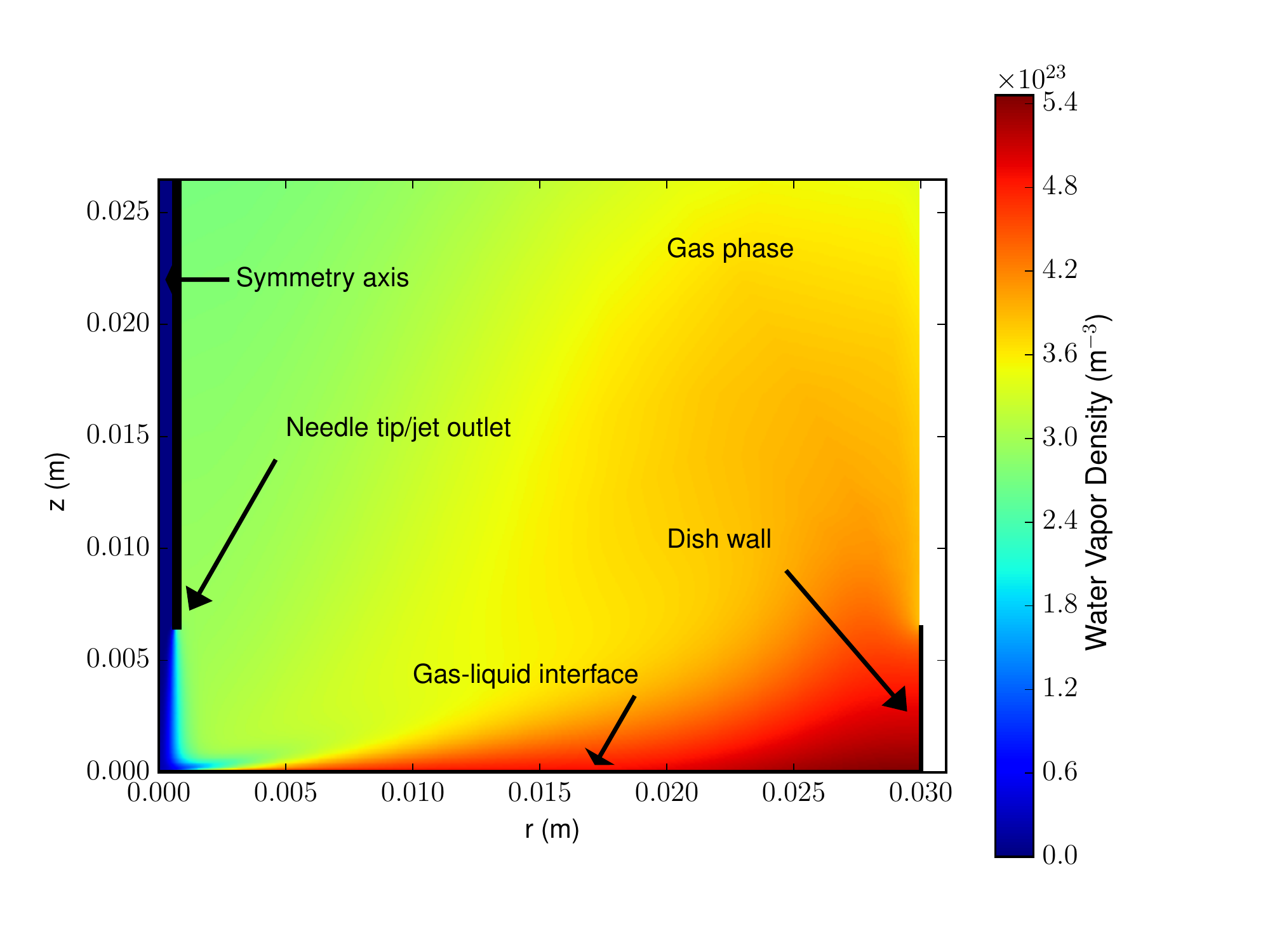}
        \caption{2-D water vapor profile at t = 1000 seconds. Red color represents higher values of water vapor concentration; blue color lower values. As implemented through Antoine's equation, water vapor concentration at the interface is highest where the temperature is highest. Role of convection in water vapor profile is evident in decreased concentration near the streamer/jet.}
        \label{fig:water_2D_profile}
\end{figure}

The temperature and water vapor profiles at the end of the simulation (t = 1000 seconds) are shown in figures \ref{fig:temp_2D_profile} and \ref{fig:water_2D_profile}. Notably, the temperature in the bulk liquid has fallen close to 10 K from its initial value of 300 K. This drop in temperature in the bulk liquid is an example of the classical wet-bulb/dry-bulb problem found in chemical engineering texts. Water vapor present in the gas above the interface is whisked away by convection. In order to maintain the equilibrium vapor pressure required by Antoine's equation, liquid water must be evaporated, consuming heat and lowering the temperature at the interface. Heat then migrates from the bulk liquid to the surface, leading to bulk liquid cooling. In this problem we have assumed that the impinging gas is at room temperature, and subsequently convection-induced evaporation leads to cooling of the bulk liquid below room ambient. However, if there is significant plasma heating of the gas, it is possible that heating of the liquid above the initial temperature will be observed experimentally. The important point is that the natural coupling between heat transport and evaporation leads to significant spatial variation in temperature, particularly at the interface, and a cooling of the bulk liquid relative to the gas. An example of the steep temperature gradients is shown in figure \ref{fig:temp_1D_profile} where the gas phase temperature changes by 10 K over the span of 200 $\mu$m, immediately above the liquid surface. Because reaction rates typically exhibit Arrhenius dependence on temperature, physical factors that introduce steep temperature gradients should be included in any model that wishes to accurately predict plasma-liquid chemistry. Though not shown here, a simulation was conducted in which temperature and water-vapor transport were de-coupled and the interfacial temperature gradient removed. Compared to the coupled case, concentrations of long-lived aqueous species like H$_2$O$_2$, NO$_2^-$, and NO$_3^-$ differed by as large as factors of two at the end of the simulation, demonstrating the importance of accounting for evaporation-induced temperature gradients.  

\begin{figure}[htb]
    \centering
        \includegraphics[width=\textwidth]{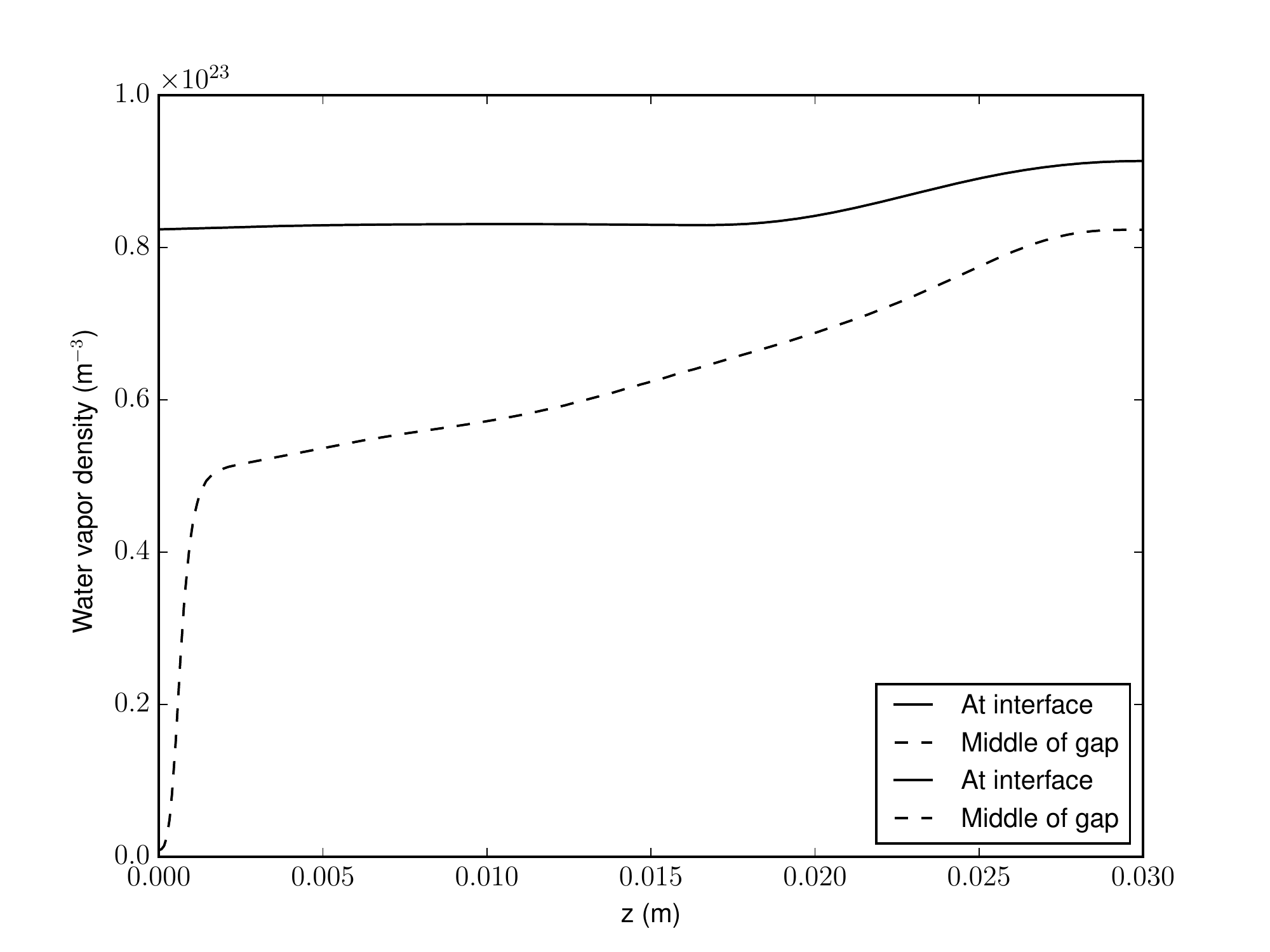}
        \caption{Radial water vapor profiles. Horizontal axis is the radial coordinate. The top curve (solid) corresponds to the water vapor concentration at the interface. The bottom curve (dashed) shows the strong water vapor radial dependence in the middle of the streamer/jet gap. Gradient is largest near and inside the discharge region.}
        \label{fig:water_1D_profile}
\end{figure}

Another instance of transport coupling that may impact plasma-liquid chemistry is the significant radial gradients in the concentration of water vapor between the needle tip/jet outlet and gas-liquid interface. An example of these gradients induced by convective forces are shown in figure \ref{fig:water_1D_profile}. Examining the dotted curve, which gives the radial distribution of water vapor half-way between the jet outlet/needle tip and liquid surface, one can see that the water vapor concentration drops precipitously in the region of the discharge (r $<$ 2 mm). In the center of the discharge (r $=$ 0) the water vapor concentration is essentially zero. This large drop in the concentration of water vapor in the active discharge region due to convection suggests that the concentrations of plasma species which rely on H$_2$O as a precursor will be reduced at increasing distances from the liquid surface, relative to  discharges where diffusion is the dominant mechanism of mass transport.

\begin{figure}[htpb]
    \centering
    \begin{subfigure}[b]{.7\textwidth}
        \includegraphics[width=\textwidth]{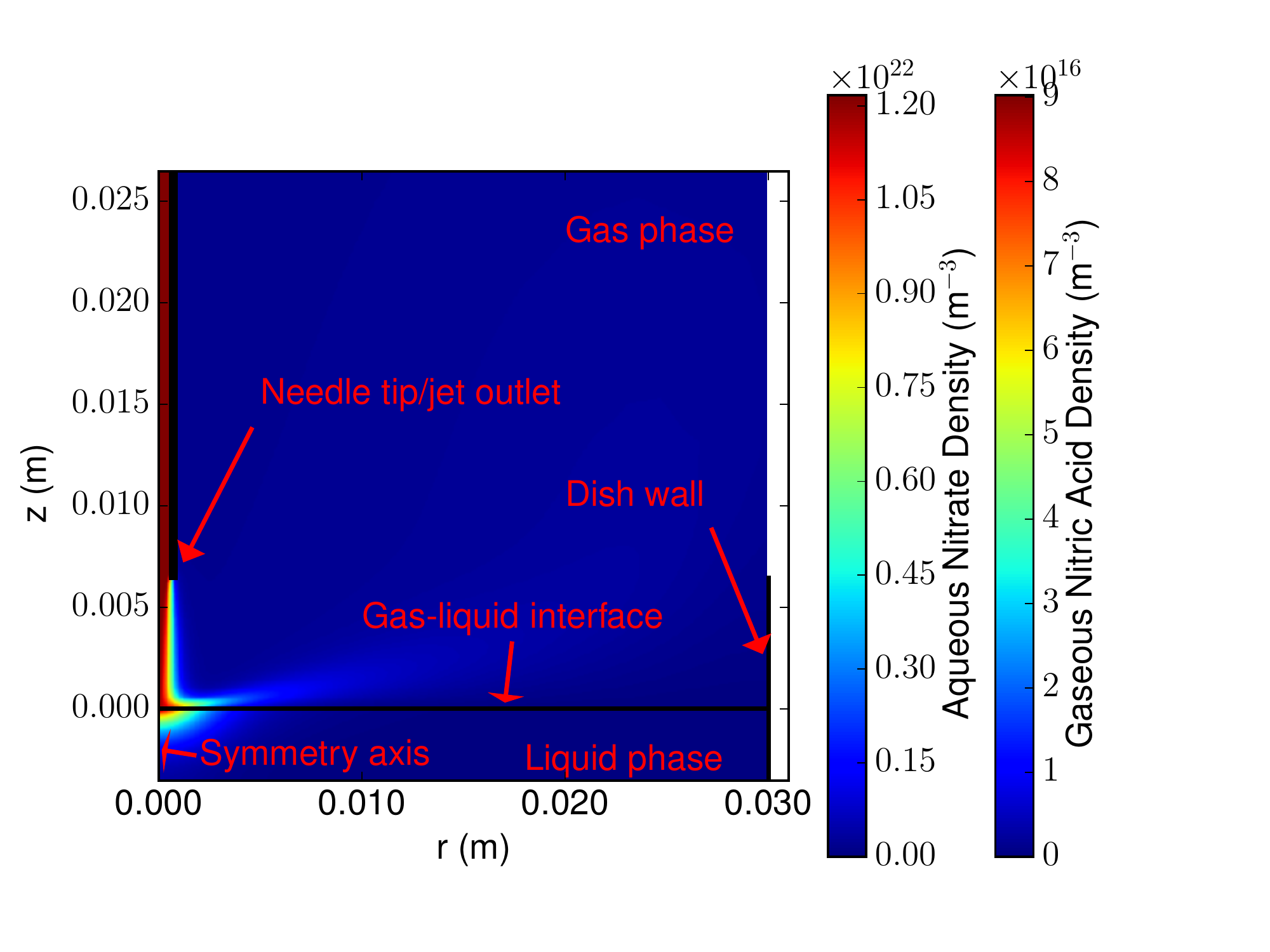}
        \caption{Spatial distribution of HNO$_3$ with liquid convection turned on.}
        \label{fig:HNO3_convec}
    \end{subfigure}
    \begin{subfigure}[b]{.7\textwidth}
        \includegraphics[width=\textwidth]{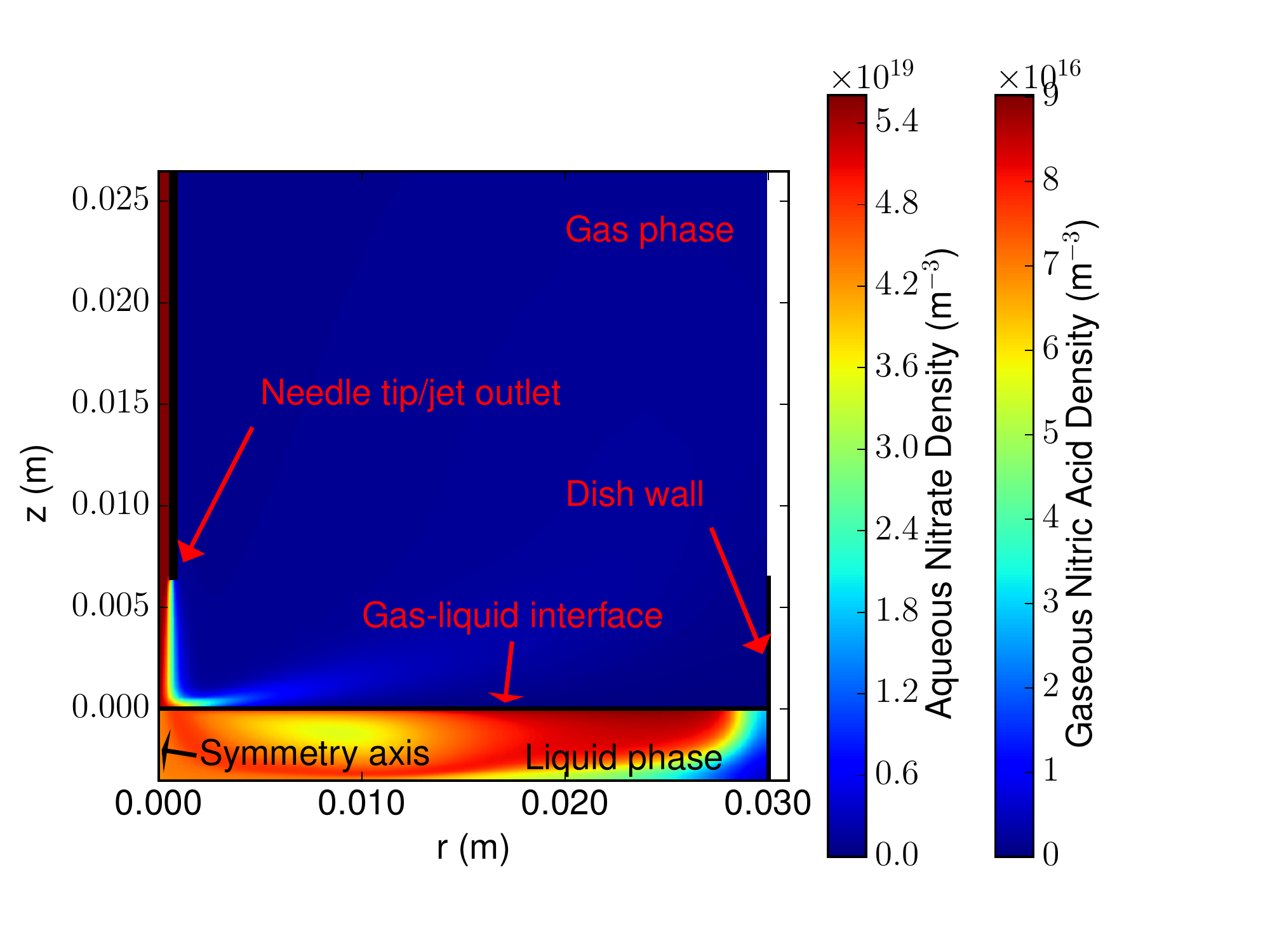}
        \caption{Spatial distribution of HNO$_3$ with liquid convection turned off.}
        \label{fig:HNO3_diffus}
    \end{subfigure}
    \caption{}
    \label{fig:HNO3}
\end{figure}

\begin{figure}[htpb]
    \centering
    \begin{subfigure}[b]{.63\textwidth}
        \includegraphics[width=\textwidth]{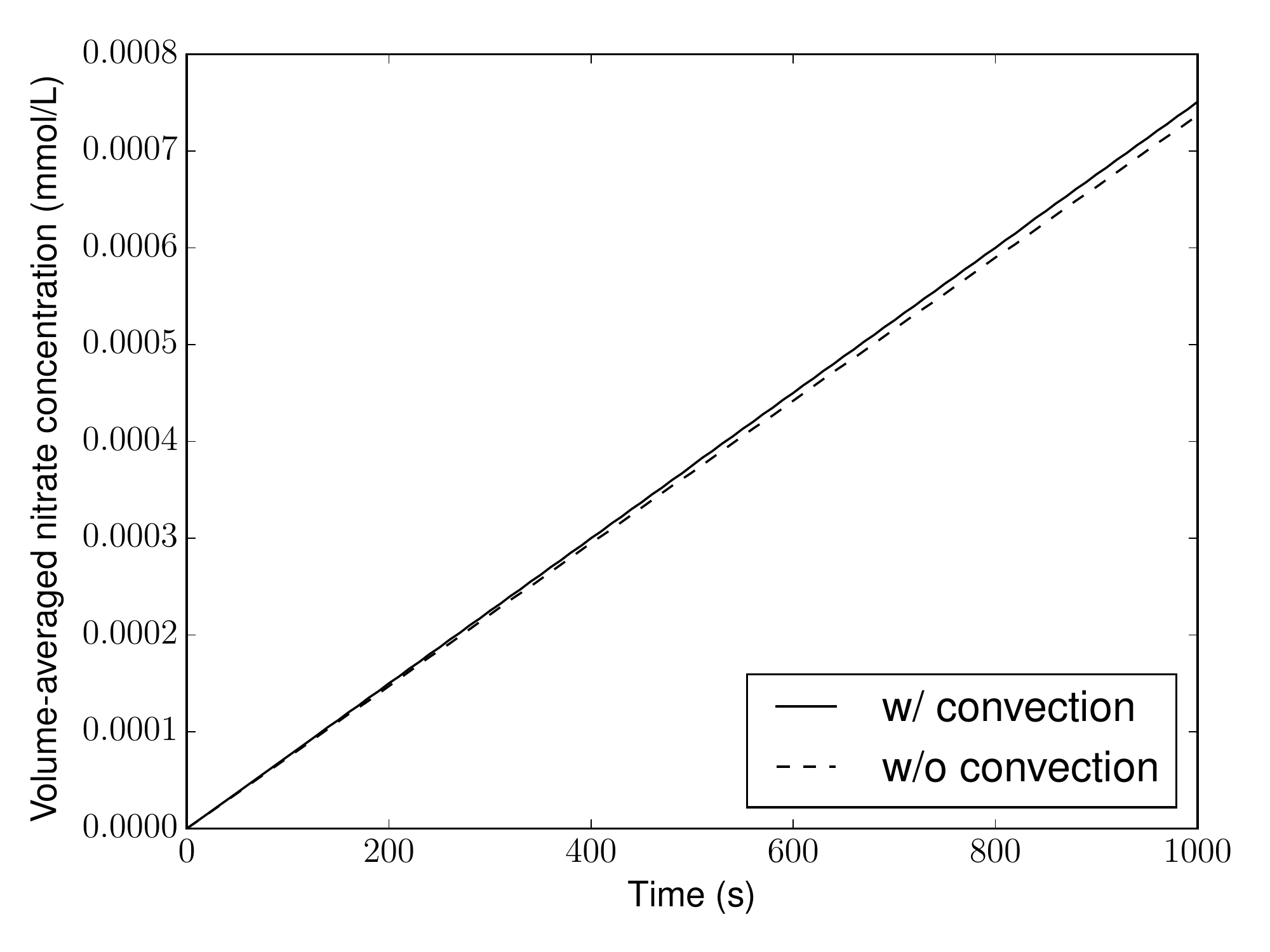}
        \caption{Comparison of volume averaged uptake of nitrate (HNO$_3$ before dissolution) with liquid convection toggled on or off. Very little difference between the two cases.}
        \label{fig:HNO3_mass_compare}
    \end{subfigure}
    \begin{subfigure}[b]{.63\textwidth}
        \includegraphics[width=\textwidth]{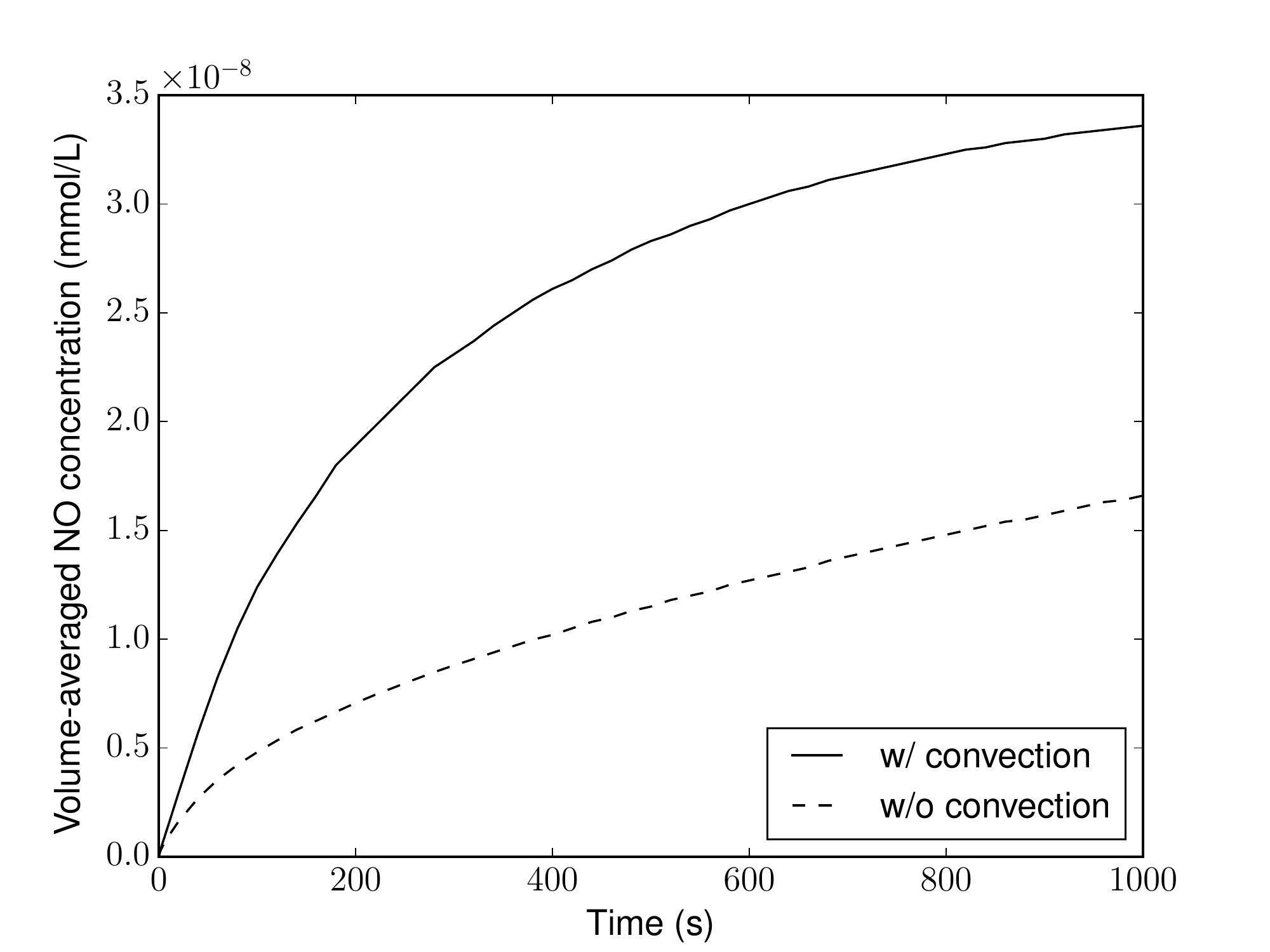}
        \caption{Comparison of volume averaged uptake of NO with liquid convection toggled on or off. The presence of convection increases the volume-averaged concentration by roughly a factor of 2 over the course of the simulation. Note that the vertical scale is much smaller for NO than for HNO$_3$ in \ref{fig:HNO3_mass_compare} because of NO's hydrophobicity}
        \label{fig:NO_mass_compare}
    \end{subfigure}
    \caption{}
    \label{fig:excel}
\end{figure}

This model also addresses the role that convection plays in dissolution rates of different gaseous species. For this analysis, reactions are turned off, reducing mass transport to only convection and diffusion. As one might intuitively expect, the induced convective flow in the liquid significantly changes the spatial distribution of aqueous species relative to a diffusion only case, as demonstrated for HNO$_3$ in figures \ref{fig:HNO3_convec} and \ref{fig:HNO3_diffus}. However, what is perhaps not intuitive is that though the HNO$_3$ spatial distribution changes dramatically depending on whether convection is present in the liquid, the volume-averaged uptake of HNO$_3$ does not change from diffusion-dominated to convection-dominated cases as shown in figure \ref{fig:HNO3_mass_compare}. If a hydrophobic specie like NO is examined instead of a hydrophilic specie like HNO$_3$, it is observed that the presence of liquid convection increases volume-averaged uptake significantly, as illustrated in figure \ref{fig:NO_mass_compare}. This fundamental difference in behavior between hydrophilic and hydrophobic species can be explained in terms of lumped mass transfer resistances. For a hydrophilic specie, the dominant resistance to interfacial transfer is in the gas-phase, whereas for a hydrophobic specie the dominant resistance to transfer occurs in the liquid phase. \cite[p. 249]{carberry2001chemical} Consequently, when convection is added to the liquid phase, effectively reducing the liquid-side mass transfer resistance, the overall resistance to mass transfer decreases and the volume-averaged uptake increases significantly for hydrophobic species like NO whereas the change in overall resistance is miniscule for hydrophilic species like HNO$_3$.

\begin{figure}[htb]
    \centering
    \includegraphics[width=\textwidth]{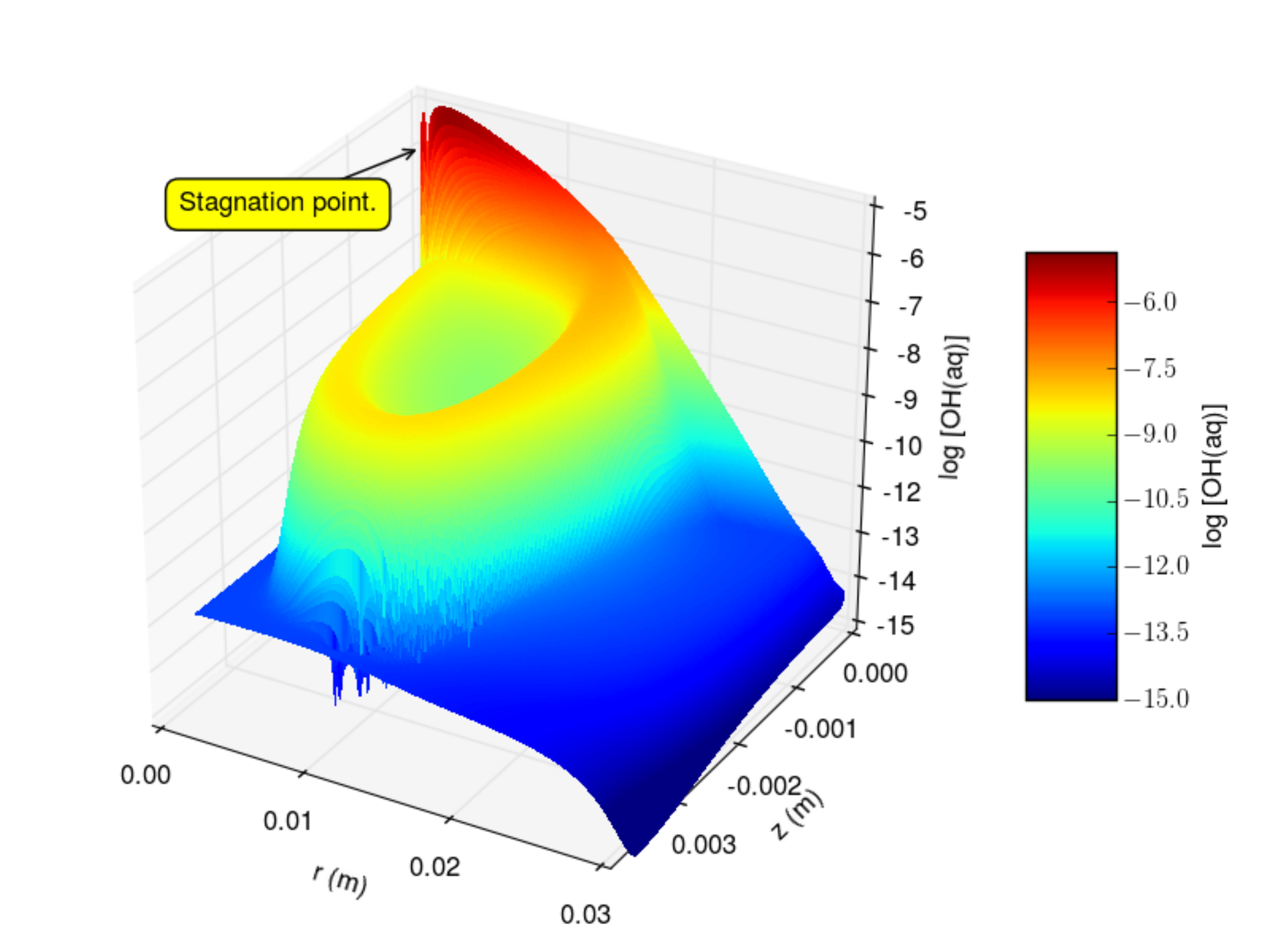}
    \caption{3D plot of log$_{10}$(OH(aq)) as a function of position for t = 1000 seconds. Large interfacial gradients are evident, particularly at the stagnation point (r=z=0). Note the effect of convection in the hole in OH concentration in the center of the plot. Some numerical noise is observed at very low OH concentrations towards the bottom of the dish. Note that the r and z axes have different scales}
    \label{fig:log_OH_conc}
\end{figure}

\begin{figure}[htb]
    \centering
    \includegraphics[width=\textwidth]{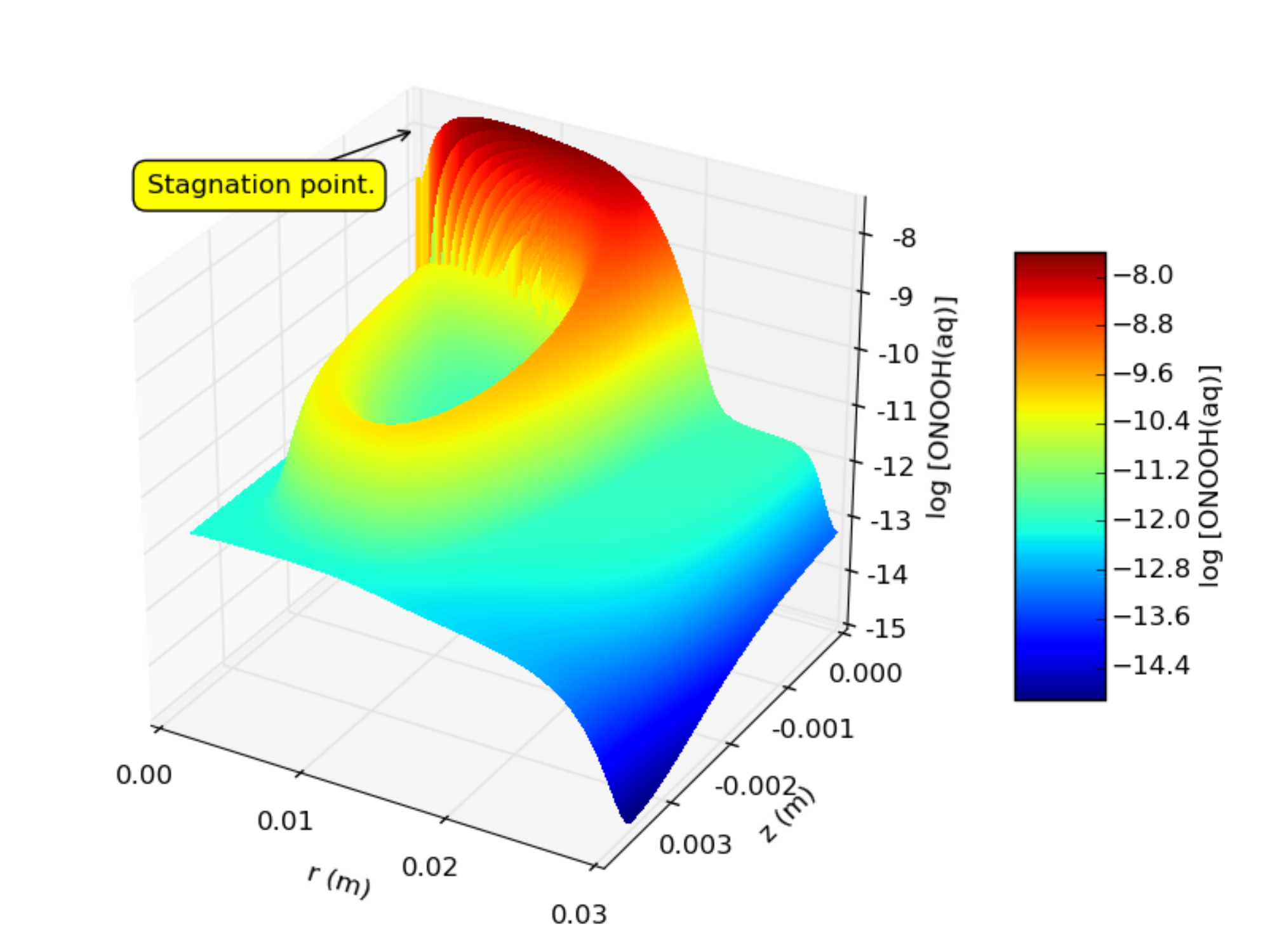}
    \caption{3D plot of log$_{10}$(ONOOH(aq)) as a function of position for t = 1000 seconds. As with OH, large interfacial gradients are evident as is the effect of the liquid convection loop. Note that the r and z axes have different scales}
    \label{fig:log_ONOOH_conc}
\end{figure}

When the full reaction set is considered, large gradients in reactive species concentrations emerge at the interface. Of particular interest for biomedical or pollutant degradation applications is the distribution of hydroxyl radical in the aqueous phase. Figure \ref{fig:log_OH_conc} shows a 3D plot of the base 10 log of OH(aq) concentration versus radial and axial position for t = 1000 seconds. Interfacial gradients in OH concentration are prominent in the region where the discharge impinges on the water surface. At the stagnation point (r=0) the OH concentration drops by approximately 9 orders of magnitude over the span of 50 $\mu$m in the axial direction. Moving away from where the streamer/jet touches the surface, the interfacial gradients become less pronounced. Even so, the largest OH concentration away from the surface and in the bulk solution is 3-4 orders of magnitude lower than the peak OH concentration which occurs at the interface and at the center of the impinging streamer/jet. The presence of the liquid phase convective loop is evident from the OH concentration hole in the center of Figure \ref{fig:log_OH_conc}. The effect is also seen in the center of Figure \ref{fig:log_ONOOH_conc} for ONOOH.

\begin{figure}[htb]
    \centering
    \includegraphics[width=\textwidth]{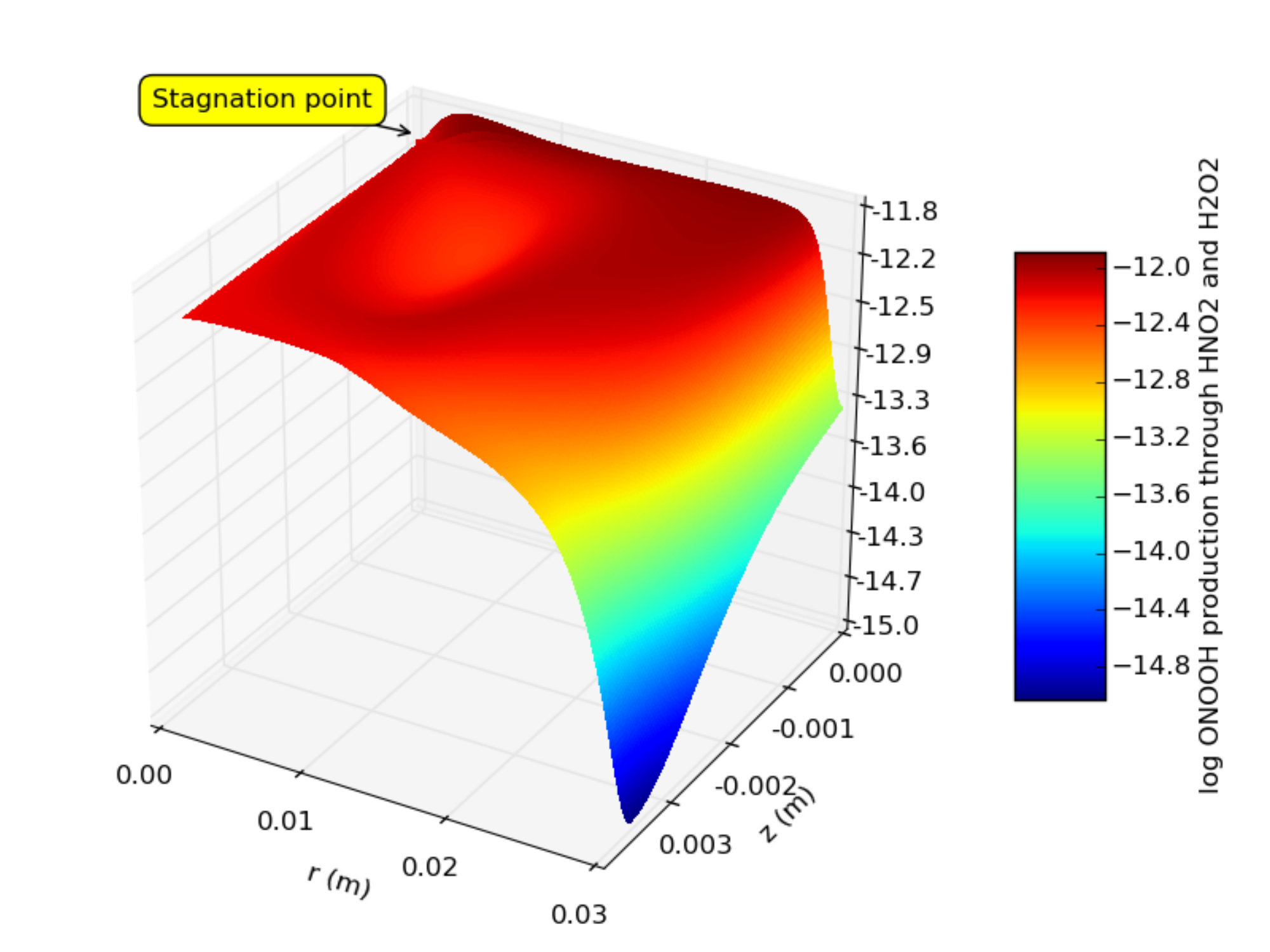}
    \caption{3D plot of the base 10 logarithm of the producation rate of ONOOH through reaction \ONOOHlong{} listed in table \ref{tab:rxns}. Because of the relative uniformity in the distribution of H$^+$, H$_2$O$_2$, and NO$_2^-$, the production of ONOOH through reaction \ONOOHlong{} is much more uniform than the overall concentration profile of ONOOH. Note that the r and z axes have different scales.}
    \label{fig:log_ONOOH_prod_rate}
\end{figure}

Another plasma-generated specie which has received much attention in the literature is peroxynitrous acid and its conjugate base peroxynitrite. In the current model, ONOOH exists only in the liquid phase and can be created through two mechanisms: reactions \ONOOHlong{} and \ONOOHshort{} in table \ref{tab:rxns}. Reaction \ONOOHlong{} was the topic of an excellent study in \cite{Lukes2014b} and is hypothesized to be a key player in the long-term anti-bacterial efficacy of plasma activated water. Figure \ref{fig:log_ONOOH_conc} shows the base 10 log of ONOOH concentration as a function of r and z in the aqueous phase. As with OH, there are large interfacial concentration gradients. At r=0, the ONOOH concentration drops by 5 orders of magnitude over 30 $\mu$m in the axial direction. Away from the stagnation point, especially where the convective current flows away from the interface, the interfacial gradient is much smaller but again as with OH, the highest bulk concentration is orders of magnitude lower than the peak surface concentration. The reason for these large ONOOH gradients is the relative dominance of reaction \ONOOHshort{} over reaction \ONOOHlong{} for the given model inputs. Over the course of the simulation, almost 7000 times more ONOOH is produced through the reaction of OH and NO$_2$ than through the reaction of H$^+$, H$_2$O$_2$, and NO$_2^-$. As shown in figure \ref{fig:log_OH_conc}, hydroxyl does not penetrate any more than a few tens of microns into the liquid phase, so consequently all ONOOH produced through OH is produced within a few tens of microns of the liquid surface. If we examine the production of ONOOH through reaction \ONOOHlong{}, it is observed to be much more uniform as illustrated in figure \ref{fig:log_ONOOH_prod_rate}. This is because of the relative uniformity in concentration of H$^+$, NO$_2^-$, and H$_2$O$_2$, although there is a peak in their concentrations and consequently in their production of ONOOH in the vicinity of the impinging streamer/jet. Once the streamer is turned off and surface fluxes of ON and NO$_2$ are removed, production of ONOOH(aq) will proceed almost exclusively through reaction \ONOOHlong{}, resulting in a mostly homogeneous distribution. This is relevent for applications of PAW since ONOOH is the long-term intermediary for production of OH through reaction \OHfromONOOH{}.

The results presented here suggest two different regimes of activity in the solution: surface and bulk. When the discharge is on, reactivity in the form of OH is confined almost entirely to the plasma-liquid interface. Bulk solution concentrations of OH are several orders of magnitude lower than the surface concentration. This is also true of other potentially biologically important and highly reactive reagents such as NO and NO$_2$ radicals. Less reactive plasma-generated RONS like H$_2$O$_2$ and NO$_2^-$ have significantly longer lifetimes and can be transported into the bulk solution where they can react and form ONOOH, a precursor to OH and NO$_2$. However, for the model inputs used here, the bulk reactivity generated through H$_2$O$_2$ and NO$_2^-$ precursors is orders of magnitude less on a per unit time scale than the surface reactivity coming from direct fluxes of plasma generated radicals. When the discharge is turned off, generation of reactive species through H$_2$O$_2$ and NO$_2^-$ will persist for a while in the bulk. The results in \cite{Lukes2014b} show a nitrite half-life of 2-3 hours in acidic solution; Traylor et. al. \cite{Traylor2011h} observed anti-bacterial efficacy of PAW, which they attributed to ONOOH and its products, for several days after plasma treatment. If one assumes that every mole of H$_2$O$_2$ present in solution at the conclusion of our simulation reacts over time to form ONOOH and 30\% of ONOOH dissociation creates OH, then the amount of OH coming from this long-term bulk reaction is 32\% less than the amount of OH that comes from surface fluxes while the ``discharge" is on. This comparison suggests that if one is willing to wait many hours or several days, bulk reactivity can approach surface reactivity on an order of magnitude scale. However, if an application demands speed and efficient utilization of plasma generated reactivity, the target must be placed right at the plasma-liquid interface. If the target is removed from the plasma by millimeters or even hundreds of $\mu$m of aqueous solution, its rate of treatment will rely on a serial process: the rate of transport of longer-lived species like H$_2$O$_2$ and NO$_2^-$ through the bulk fluid followed by the rate of formation of peroxynitrous acid chemistry.

\begin{figure}[htb]
    \centering
    \includegraphics[width=\textwidth]{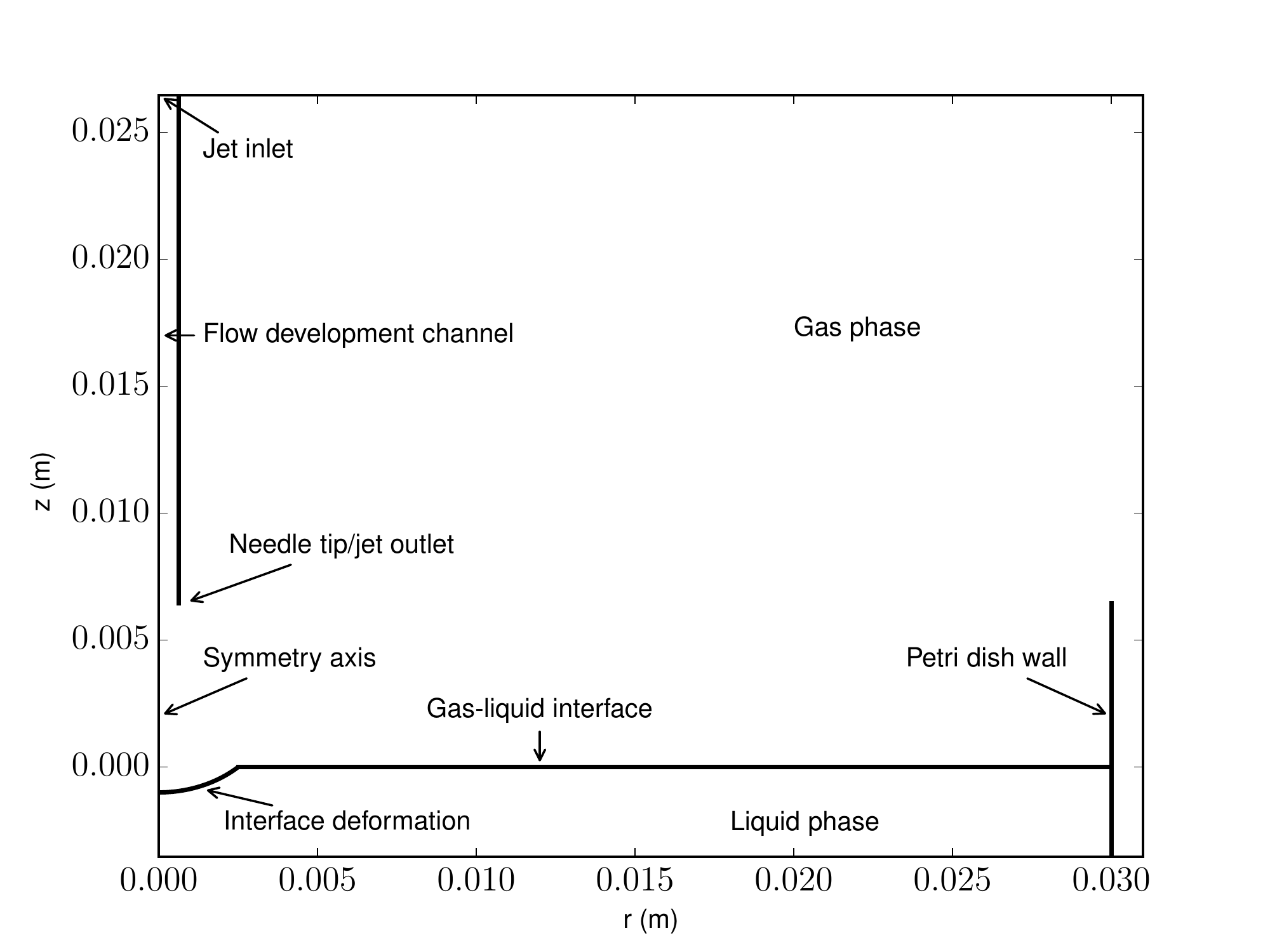}
    \caption{Geometry for deformed interface simulations.}
    \label{fig:deformed_geom}
\end{figure}

\begin{figure}[htb]
    \centering
    \includegraphics[width=\textwidth]{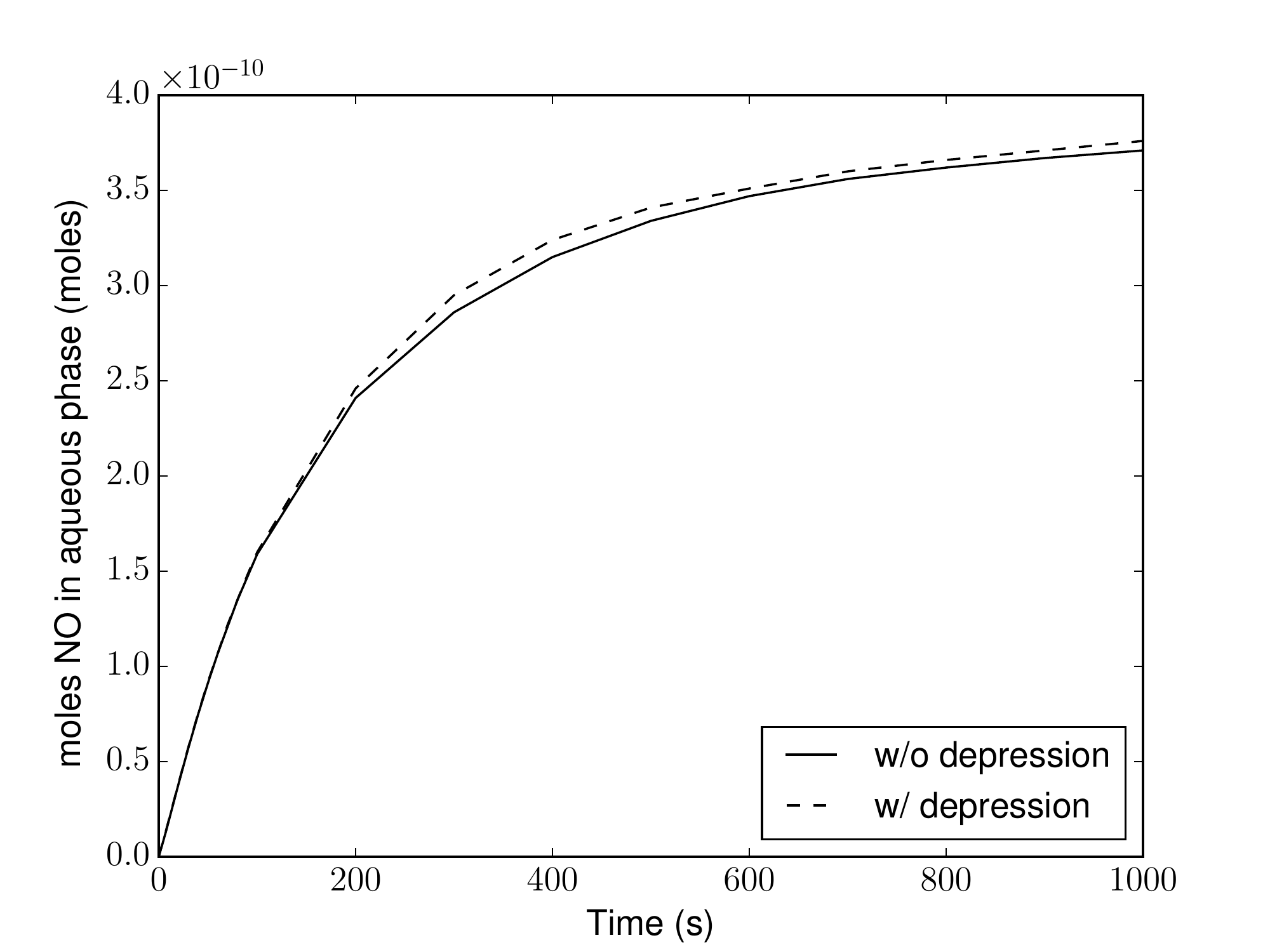}
    \caption{Comparison of total NO(aq) uptake as a function of time for cases in which interface deformation is included and not included. As shown in the figure, the interface deformation has very little effect on the macrosocpic NO uptake. The effect for hydrophilic species is expected to be even less}
    \label{fig:NO_deform_compare}
\end{figure}

Before concluding, it is worth touching on the assumption of a flat interface. This work discusses the role gas phase convection plays in determining momentum, heat, and mass transport across a gas-liquid interface and into the bulk liquid. Species uptake is used as the criteria for determining whether to include self-consistent deformation of the interface by the gas flow. To make this determination, the shape of the interface deformation was observed experimentally and the deformation was introduced into the model geometry as shown in figure \ref{fig:deformed_geom}. The depth of the deformation was supported by a calculation balancing gravitational ($\rho$gh) and convective ($\frac{1}{2}\rho$v$^2$) stresses. The fluid flow simulation was then run until it reached stead-state, and then the heat and mass transfer equations were solved. In this way, though the interface deformation was not self-consistently determined, its influence on variables of interest could be analyzed. Shown in \ref{fig:NO_deform_compare} is the effect of the interface deformation on total solution uptake of NO(aq). As can be seen in the figure, the effect is minute. Because of the characteristics shown in Figure \ref{fig:excel}, the effect of the interface deformation on hydrophilic species transport is expected to be even less. Subsequently, for the purpose of this work, the interface deformation can be safely neglected.  

\section{Conclusion}

A qualitative study of the momentum, heat, and mass transport in a convective plasma-liquid system has been conducted. Several interesting results were found. Convective flow of water vapor away from the interface leads to a sharp temperature gradient between the bulk gas and liquid phases. This sharp gradient is important for accurately determining temperature-dependent rate coefficients in both the gas and liquid phases. Additionally, convection drives water vapor away from the discharge region except immediately above the liquid surface; this could have important consequences for gas-phase generation of reactive species that depend on water as a precursor. Induced convection in the liquid phase substantially changes the spatial distribution of aqueous species and increases the volume-averaged uptake of hydrophobic species but interestingly has little effect on the volume-averaged uptake of hydrophilic species. This phenomena occurs because the majority of resistance to interfacial transfer is in the gas phase for hydrophilic species and in the liquid phase for hydrophobic species; consequently, decreasing the liquid-phase mass transfer resistance by adding liquid convection has a significant impact for hydrophobic but not for hydrophilic molecules.

Perhaps the most interesting result of the study is the sharp distinction between reactivity at the surface and in the liquid bulk. Though the limited penetration (tens of $\mu$m or less) of reactive neutral radicals is well publicized in prominent biochemistry texts like \cite{Halliwell}, this phenomena has yet to receive significant attention in the plasma-liquid literature with the exception of \cite{Chen2014a}. Concentrations of species of interest for plasma-medicine and other applications (e.g. OH, NO, NO$_2$, ONOOH) fall by as many as 9 orders of magnitude within 50 $\mu$m of the interface. In a relatively pure aqueous solution as is modeled here, the process responsible for conveying plasma reactivity to a target may be transport of H$_2$O$_2$ and acidified nitrite followed by a close-proximity reaction to form ONOOH and then OH and NO$_2$. For targets in aqueous biological systems, the conveyors of reactivity may be proteins modified by primary plasma species. What can be stated from the results presented here and in \cite{Chen2014a} is that these conveyors of plasma reactivity are almost probably not neutral radicals generated directly by the plasma; rather they are likely secondary species or more stable compounds like H$_2$O$_2$ and nitrite.    

\FloatBarrier

\bibliographystyle{unsrt}
\bibliography{paper}
\end{document}